\documentclass{svproc}
\usepackage{graphicx}
\usepackage{graphics}
\usepackage{amsmath,amsfonts}
\usepackage{xr}
\usepackage{url}
\usepackage{float}

\usepackage{multicol}
\usepackage{lipsum}
\usepackage{mwe}
\usepackage{subfigure}
\usepackage{hyperref}
\usepackage[table,xcdraw]{xcolor}
\usepackage{lscape}
\usepackage{longtable}
\usepackage{nomencl}
\makenomenclature
\usepackage{colortbl}
\hypersetup{
    colorlinks=true,
    linkcolor=blue,
    filecolor=blue,      
    urlcolor=blue,
    citecolor = blue,
    pdftitle={NODYArXiv},
    pdfpagemode = {FullScreen}}
\usepackage[nameinlink,noabbrev,capitalize]{cleveref}

\begin{document}
\mainmatter              % start of a contribution
\title{A Set-valued Impact Law Approach for Modeling and Analysis of Rigid Contact Universal Joint with Clearance}
\titlerunning{Multibody Systems}  % abbreviated title (for running head)
\author{Junaid Ali\inst{1}, Gregory Shaver\inst{1},
Anil K. Bajaj\inst{1}}
\authorrunning{Junaid Ali et al.} % abbreviated author list (for running head)
\tocauthor{Junaid Ali, Gregory Shaver, Anil K. Bajaj}
\institute{Ray W. Herrick Laboratory, Purdue University, West Lafayette, IN 47907, USA,\\
\email{ali181@purdue.edu}}

\maketitle              % typeset the title of the contribution

\begin{abstract}

This study presents a dynamic model of a universal joint (U-Joint) with radial clearance, focusing on the rigid unilateral frictional contacts at the crosspiece and yoke interfaces. Unlike previous models that neglect crosspiece inertia and interface friction, this work incorporates these effects using a set-valued impact law based on Signorini's condition with Coulomb friction, capturing the complex non-smooth dynamics introduced by radial clearance. Numerical simulations of a 2 degrees-of-freedom (DOF) shaft system reveal the critical influence of clearance on U-Joint dynamic behavior, including impact-induced oscillations, quasi-periodic motion, and chaotic dynamics, which are essential for accurate driveline modeling and real-time control in automotive, aerospace, and precision medical applications.

\end{abstract}

\section{Introduction} \label{intro}

Mechanical clearances, while essential for enabling relative motion between interconnected components, introduce impacts characterized by abrupt velocity changes and impulsive contact forces in multibody systems \cite{Peterka1999}. The choice between rigid and deformable contact models significantly influences the accuracy and computational efficiency of impact simulations. Rigid models, which assume instantaneous energy transfer, are computationally efficient but overlook the deformation present in real-world contacts, whereas deformable models capture the continuous nature of contact forces and energy dissipation over a finite duration, providing higher fidelity at increased computational cost \cite{Flores2023,Tian2018}. For instance, ball and roller bearings typically employ deformable contact models due to the relative hardness difference between the balls and raceway surfaces, allowing for minor elastic deformation and penetration. However, rigid models remain preferable for high-stiffness, impact-dominated systems where instantaneous energy exchange is a more accurate assumption. For rigid multibody systems, the set-valued impact modeling approach for rigid contacts, as detailed in \cite{Leine2004,Glocker2001}, provides a robust framework for analyzing instantaneous energy transfer during impacts under the impenetrability assumption.  \par

\begin{figure}[ht]
    \centering
    \includegraphics[width = 0.8\textwidth]{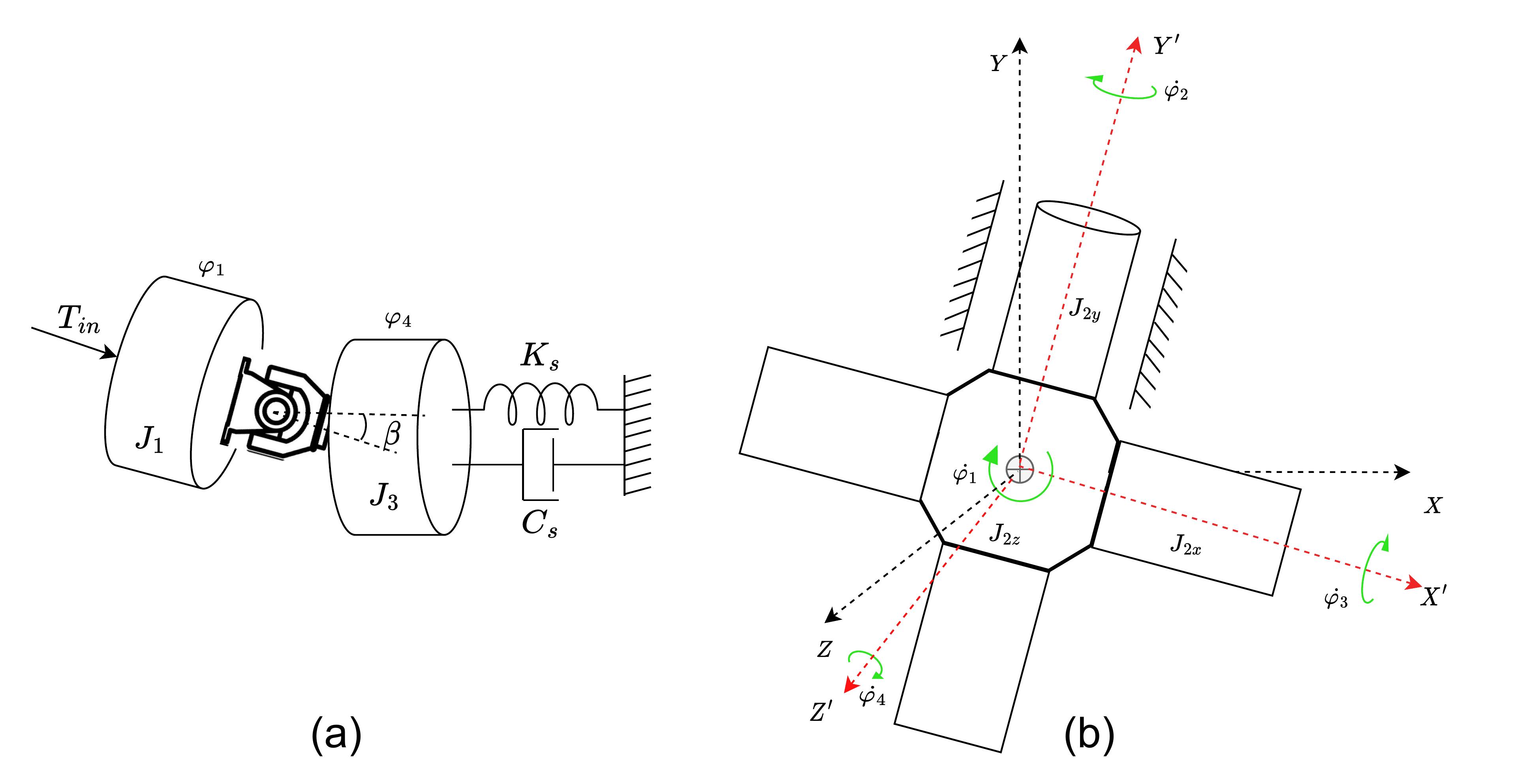}
    \caption{(a) Schematic representation of a 2-DOF shaft system interconnected with U-Joint misaligned at $\mathrm{\beta}$ angle, and (b) Illustration of crosspiece with respect to input yoke in rotational frame of reference $(X',Y', Z')$ and inertial frame of reference $(X,Y,Z)$. }
    \label{fig:fig1}
\end{figure}

\begin{figure}[ht]
    \centering
    \includegraphics[width=0.65\textwidth]{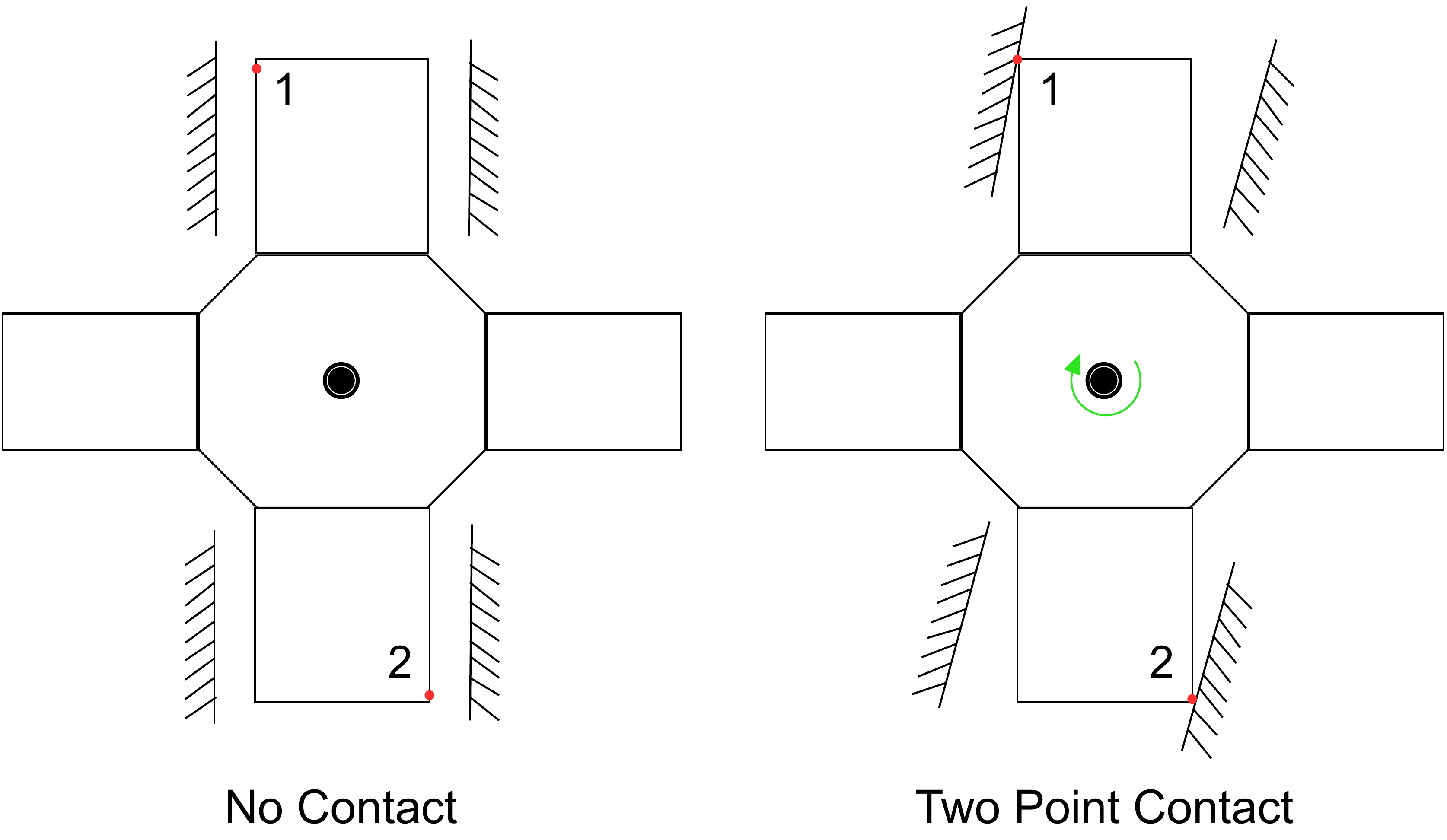}
\caption{Illustration of possible contact points (red dots) between input yoke and crosspiece arm.}

    \label{fig:fig2}
\end{figure}

\begin{figure}[ht]
    \centering
    \subfigure[]{\includegraphics[width=0.35\textwidth]{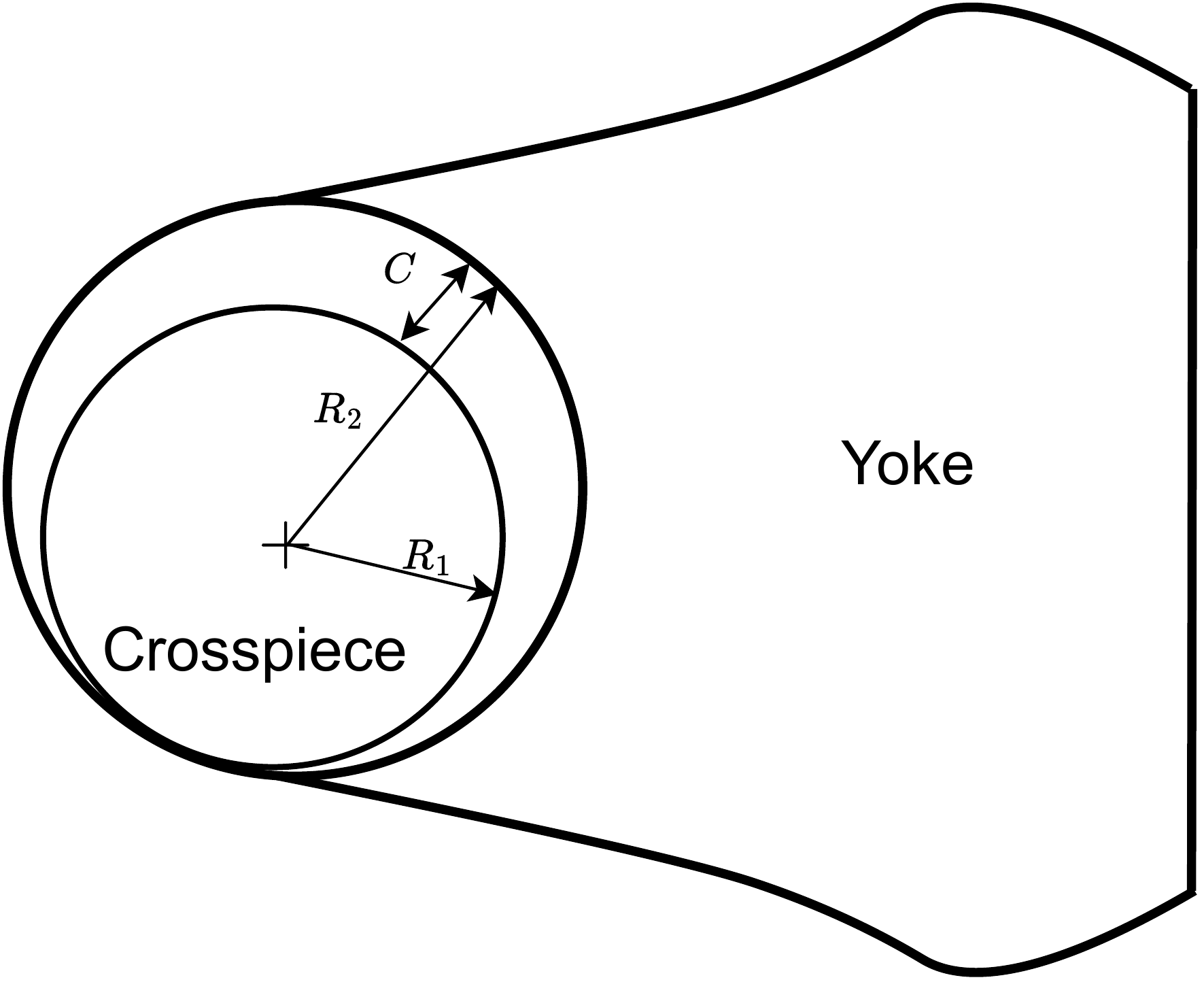}}
    \subfigure[]{\includegraphics[width=0.35\textwidth]{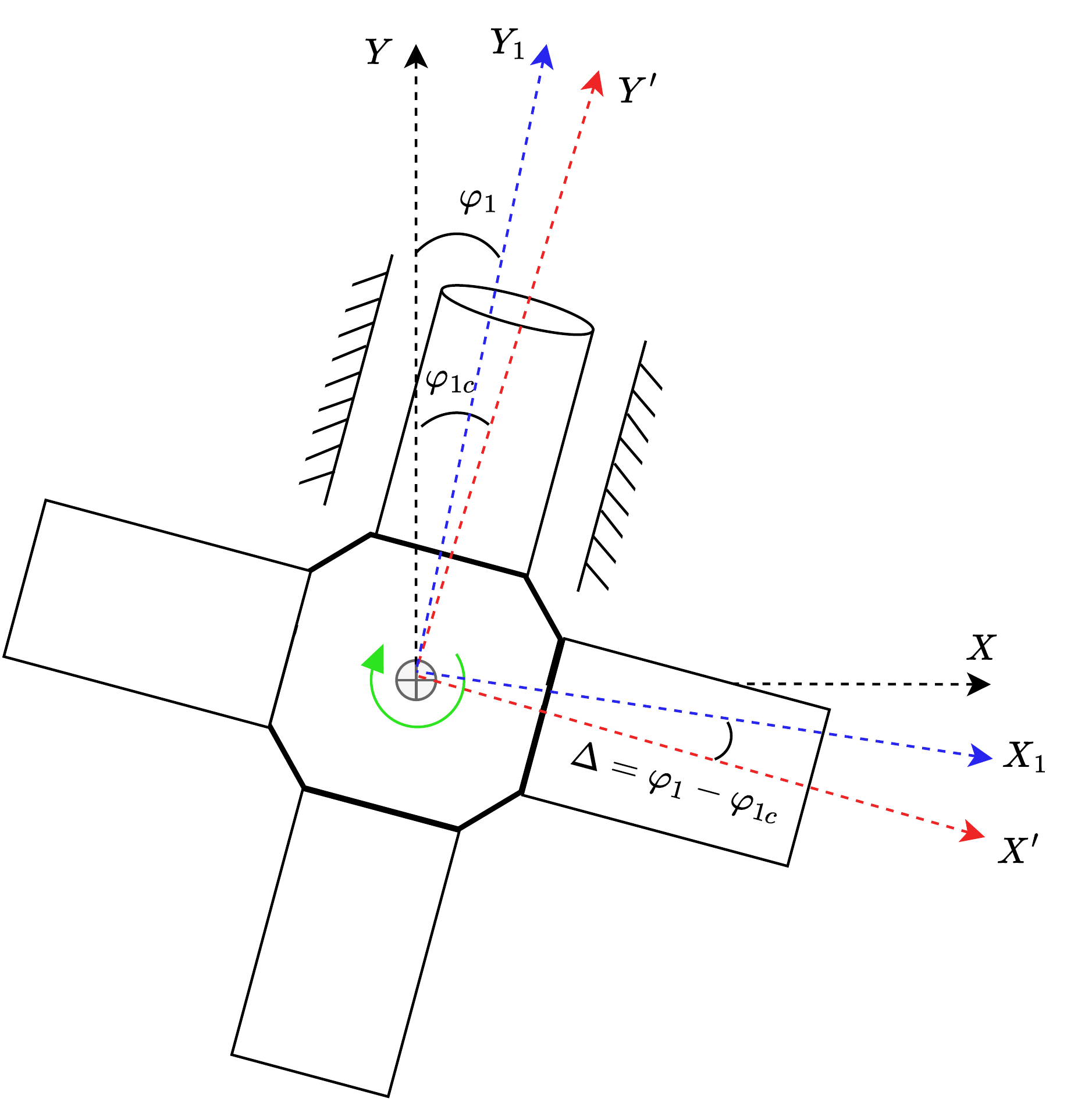}}
\caption{(a) Illustration of clearance \(C\) between yoke inner radius \(R_2\) and crosspiece cap outer radius \(R_1\), and (b) relative angular position of crosspiece \(\varphi_{1c}\) in the rotational frame \((X',Y')\) in red, and angular position of input yoke \(\varphi_1\) in the rotational frame \((X_1,Y_1)\) with respect to the fixed frame \((X,Y)\).}
    \label{fig:fig3}
\end{figure}

Despite significant progress in contact dynamics research, studies on U-Joint systems with radial clearances remain limited. Traditional models, such as those in \cite{Ali2024,Jianwei2013}, often employ piecewise-affine Hertzian contact force formulations, which, while effective for capturing continuous deformations, struggle to accurately capture the rapid, impulsive nature of rigid impacts due to their single-valued nature \cite{Flores2023}. These models typically neglect critical effects such as post-impact velocity recovery and the complex frictional interactions at the yoke-crosspiece interface. Moreover, while some recent work \cite{Jianwei2013} has included crosspiece inertia, it often simplifies the dynamics by omitting the nonlinear contributions from each crosspiece arm and ignoring post-impact kinematics, which are essential for accurately capturing the non-smooth behavior inherent to U-Joints with clearance. \par

To date, no comprehensive theoretical model fully captures the dynamics of U-Joints with clearances, accounting for crosspiece inertia and frictional interactions at the yoke interface during impact events. These effects can induce significant torque fluctuations, chatter, and mechanical failure in driveline systems, which are critical in automotive, aerospace, and precision medical applications. In this study, a mathematical model of a U-Joint with clearance is developed using a set-valued impact law to capture the rigid impulsive contact forces in a 2-DOF shaft system. The system's dynamics, both with and without clearance, are analyzed in the results and discussion section of this manuscript. 

\section{System Modeling} \label{sysmodel}

In this study, a 2-DOF shaft system is modeled, where the input shaft with inertia $\mathrm{J_1}$ is connected to the output shaft with inertia $\mathrm{J_{3}}$ through a U-Joint inclined at angle $\beta$ (see Figure \ref{fig:fig1} (a)). The output shaft has stiffness $\mathrm{K_s}$ and velocity-proportional damping with coefficient $\mathrm{C_s}$. The U-Joint crosspiece has polar moments of inertia $\mathrm{J_{2x}}$, $\mathrm{J_{2y}}$, and $\mathrm{J_{2z}}$ along the $\mathrm{X'}$, $\mathrm{Y'}$, and $\mathrm{Z'}$ axes, respectively (see Figure \ref{fig:fig1} (b)). The system is driven by a sinusoidal input torque $\mathrm{T_{in}(t)} = \mathrm{T_0} \sin(\mathrm{\Omega} t)$, where $\mathrm{\Omega}$ is the forcing frequency and $\mathrm{T_0}$ is the amplitude. For simplicity, only rotational clearance \cite{hummel2} is assumed between the input yoke and the crosspiece, while the output yoke and crosspiece are modeled as an interference fit \cite{Ali2024}. This reduces the system to a two-contact-point impact configuration (see Figure \ref{fig:fig2}). During contact, the lag in motion transmission causes impulsive forces at Points 1 and 2, located at the ends of the crosspiece arms in contact with the input yoke. \par

 The rotational clearance arises from the radial clearance between the inner diameter of the yoke and the outer diameter of the crosspiece cap $\mathrm{C} = R_2 - R_1$(see Figure \ref{fig:fig3} (a)). The DOF of the crosspiece is indicated by $\{\varphi_{1c}, \dot{\varphi}_{1c}, \varphi_2, \dot{\varphi}_2, \varphi_3, \dot{\varphi}_3\}$, where $\varphi_{1c}$ represents the motion of crosspiece along the $\mathrm{Z'}$-axis, $\varphi_2$ represents the relative angular rotation of crosspiece with respect to input shaft along the $\mathrm{Y'}$ axis, and $\varphi_3$ represents the relative angular rotation of crosspiece with respect to output shaft along the $\mathrm{X'}$ axis in rotational frame (see Figure \ref{fig:fig1} (b)). The DOF of the input and output shaft is indicated by the state variables $\{\varphi_1, \dot{\varphi}_1\}$ and $\{\varphi_4, \dot{\varphi}_4\}$ (see Figure \ref{fig:fig1} (b)). As shown in Figure \ref{fig:fig3} (b), the rotational clearance leads to transmission error $(\mathrm{\Delta} = \varphi_1 - \varphi_{1c})$, when both bodies are in contact $\varphi_1 = \varphi_{1c}$. This is significant because when there is no contact between the input yoke and the crosspiece, the conventional holonomic constraint between the input and output of the U-Joint will not be applicable between $\varphi_1$ and $\varphi_4$, rather it will be true for $\varphi_{1c}$ and $\varphi_4$ (see \eqref{eq:holonomic}), since there is no clearance between the crosspiece and the output yoke. The geometric relationship between the input and output rotational positions of the U-Joint inclined at an angle \(\mathrm{\beta}\) with respect to the horizontal plane, poses a holonomic constraint \eqref{eq:holonomic},

\begin{equation}
    \varphi_4 = \arctan(\tan(\varphi_{1c})/\cos(\mathrm{\beta}))
    \label{eq:holonomic}  
\end{equation}

\noindent Differentiating \eqref{eq:holonomic} with respect to time results in the rotational velocities of the crosspiece along $\mathrm{Z'}$-axis and output shaft which can be written as \eqref{eq:velocityconstraint}, 
\begin{equation}
    \Dot{\varphi}_4 = \frac{\cos(\mathrm{\beta})}{1 - \sin(\mathrm{\beta})^2\cos(\varphi_{1c})^2} \Dot{\varphi}_{1c} \equiv \eta \Dot{\varphi}_{1c}
    \label{eq:velocityconstraint}  
\end{equation}

\noindent where $\eta$ is a time-varying function strictly dependent on $\varphi_{1c}$ and $\mathrm{\beta}$. Similarly, the angular velocity relationship between crosspiece $\mathrm{Y'}$-axis and input yoke can be written as $\dot{\varphi}_2 = \nu \cdot \Dot{\varphi}_{1c}$, where $\nu = -\frac{\sin(\mathrm{\beta}) \cos(\mathrm{\beta}) \cos(\varphi_{1c})}{1 - \sin^2(\mathrm{\beta}) \cos^2(\varphi_{1c})}.$

Using Lagrange's equation of motion, the following system of equations can be used to describe the non-smooth dynamics of the U-Joint with clearance \eqref{eq:model}. The presence of clearance in the system causes the corresponding degrees of freedom of the input shaft and the crosspiece to become decoupled,

\begin{align}
\resizebox{0.85\textwidth}{!}{$
\begin{aligned}
&\begin{bmatrix}
    \mathrm{J_1} & 0 \\
    0 & \mathrm{J_{3}} \eta^2 + \mathrm{J_{2y}} \nu^2 + \mathrm{J_{2x}} \cos^2(\varphi_2) + \mathrm{J_{2z}} \sin^2(\varphi_2)
\end{bmatrix}
\begin{bmatrix}
    \Dot{\varphi}_1 \\
    \Dot{\varphi}_{1c}
\end{bmatrix} \\
&\quad +
\begin{bmatrix}
    -\mathrm{T_{in}}(t) \\
    \left(
        \mathrm{J_{3}} \eta \eta' + \mathrm{J_{2y}} \nu \nu'
        + \left( \frac{\partial \varphi_2}{\partial \varphi_{1c}} - 2\nu \right) 
        \frac{(\mathrm{J_{2x}} - \mathrm{J_{2z}})}{2} \sin(2\varphi_2)
    \right) \dot{\varphi}_{1c}^2
    + \mathrm{K_s} \eta \varphi_4
    + \mathrm{C_s} \eta \dot{\varphi}_4
\end{bmatrix}
= 0
\end{aligned}
$}
\label{eq:model}
\end{align}

\begin{figure}[ht]
    \centering
    \subfigure[]{\includegraphics[width=0.30\textwidth]{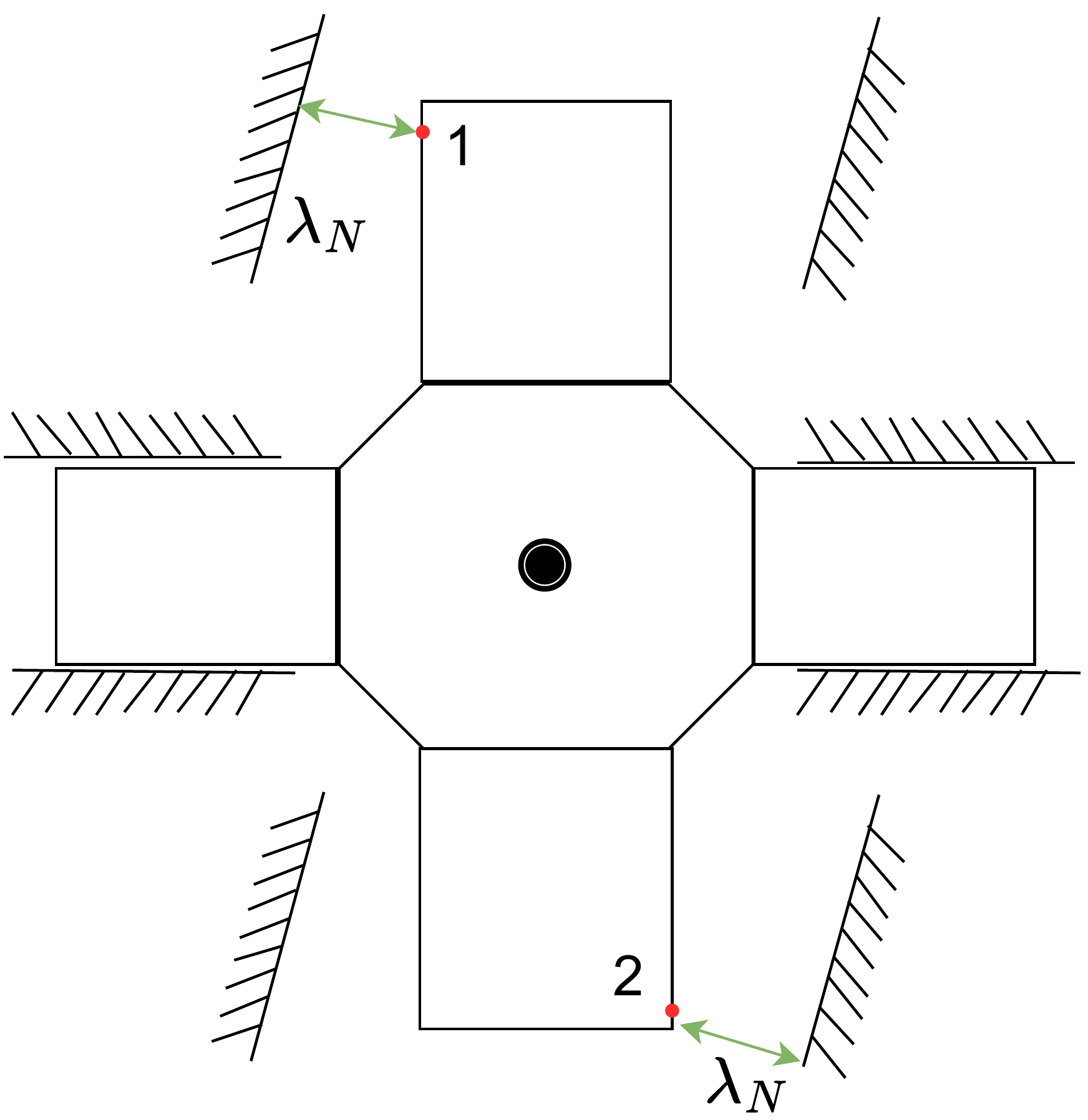}}
        \subfigure[]{\includegraphics[width=0.35\textwidth]{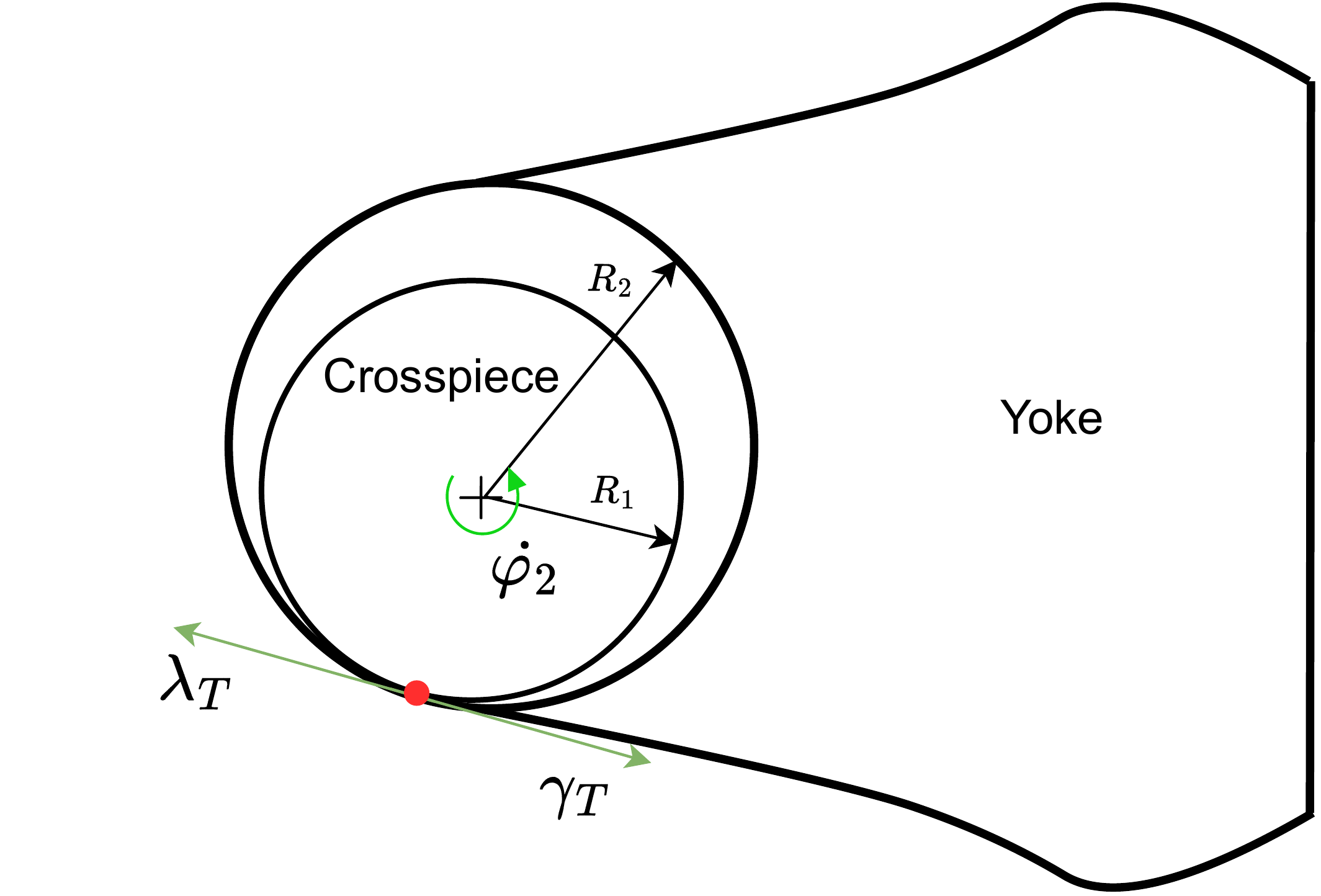}}
    \caption{(a) Relative normal contact force \(\lambda_N\) at contact points 1 and 2 (red dots) during impact, and (b) top view of input yoke and crosspiece with clearance \(C\), illustrating tangential drift \(\gamma_T\) due to frictional forces \(\lambda_T\).
}
    
    \label{fig:fig4}
\end{figure}

\noindent Further, \eqref{eq:model} can be written in condensed matrix form as shown in \eqref{eq:non_impact},
\begin{equation}
    \begin{aligned}
    \mathrm{M(t,q)}\Dot{u} - \mathrm{h(t,u,q)} = 0
    \\
    \Dot{q} = u
    \end{aligned}
    \label{eq:non_impact}
\end{equation}
where $\mathrm{M(t,q)}$ is state dependent positive definite and symmetric inertia matrix, $\mathrm{h(t,u,q)}$ is a vector of all external and gyroscopic forces acting on the system, generalized position vector $q \in \{\varphi_1, \varphi_{1c}\}$ and generalized velocity vector $u \in \{\dot{\varphi}_{1}, \dot{\varphi}_{1c}\}$. The $\eta'$ and $\nu'$ are the first derivatives of $\eta$ and $\nu$ with respect to $\varphi_{1c}$. Notice, in \eqref{eq:non_impact} the reaction forces due to impact do not appear. The reaction forces in normal and tangential direction can be added as Lagrange's multipliers, which is discussed in next section. 

\subsection{Set Valued Contact Law}
The contact event is described here as the moment when the gap between yoke and crosspiece becomes zero i.e. $\varphi_1 = \varphi_{1c}$. Due to unilateral nature of contact, the penetration or deformation is not permissible at the yoke interface. There are two possible scenarios for contact that needs to be considered: (i) the contact between left wall of the yoke and crosspiece, and (ii) the contact between right wall of the yoke and crosspiece. The mathematical equations of both gap functions are given in \eqref{eq:gapfunctions}, 

\begin{equation}
\begin{aligned}
\text{Left Wall: } \quad g_{N}^- &= -\mathrm{\Delta} \cdot \mathrm{L} + \mathrm{C} \\
\text{Right Wall: } \quad g_{N}^+ &= \mathrm{\Delta} \cdot \mathrm{L} + \mathrm{C}
\end{aligned}
\label{eq:gapfunctions}
\end{equation}

The relative normal gap functions and the events can be described using so-called Signorini's conditions \eqref{eq:signorinicond} (for more details on Signorini's condition for unilateral constraints refer to \cite{Jean1999}). When $g_{N} = 0$ implies contacts are closed (in either direction), resulting in a normal contact force $\lambda_N \geq 0$. Also, for $g_{N} > 0$ implies the contacts are open and resulting in a normal contact force $\lambda_N = 0$ (see Figure \ref{fig:fig4} (a)),

\begin{equation}
    g_N = 0, \lambda_N \geq 0 \quad \text{and} \quad   g_N > 0, \lambda_N = 0 \implies
     g_N \cdot \lambda_N = 0 
     \label{eq:signorinicond}
\end{equation}
Therefore, the normal contact force $\lambda_N$ can be represented as convex hull of all permissible contact points $\mathrm{H_M}$ \eqref{eqa:closedcontactset}, 
\begin{equation}
    \mathrm{H_M} = \{ i \in \Re | g_N \leq 0\}
    \label{eqa:closedcontactset}
\end{equation}
% \noindent The resulting contact force as a normal cone of $\mathrm{h_M}$, admissible contact point.
% \begin{equation}
%     -\lambda_{N} \in N_{\mathrm{h_M}}(g_{N}) 
%     \label{eq:normalforcecone}
% \end{equation}
The direction of generalized relative normal force would be $\mathrm{W_N} = [\frac{\partial g_N}{\partial q}]^T$.

\subsection{Set-Valued Friction Law}

When input yoke is approaching the crosspiece, the gap $g_N \rightarrow 0$. The rate at which it will be reduced will be determined by the relative normal $\gamma_N^{\pm} = \dot{g_N}^{\pm}$ velocity. The tangential slip velocity between crosspiece and yoke is given by $(\gamma_T = \nu \dot{\varphi}_1 \cdot \mathrm{R_1})$, causes frictional drag due to metal-to-metal contact interface. When contact occurs $g_N = 0$, but the tangential velocity is $\gamma_T \neq 0$, because $\dot{\varphi}_2 \neq 0$. There are two possible scenarios, the first is $\gamma_T = 0$, which represents the sticking or pure rolling due to friction $\mu$ between bodies, and the second is $\gamma_T \neq 0$, which is sliding. The crosspiece undergoes rotational motion about the $\mathrm{Y'}$-axis (green arrow, see Figure \ref{fig:fig4} (b)), which may induce stick-slip behavior at the contact points. Therefore, the set-valued description of the tangential contact forces due to friction is given as \eqref{eq:frictionsetlaw}, 
\begin{equation}
\begin{aligned}
 \texttt{Sliding:}  \quad     \lambda_{T} \in \mu \lambda_{N} \mathrm{Sgn}(\gamma_T)  \quad \text{such that} \quad  \gamma_{T} \neq 0 
 \\
 \texttt{Sticking:}  \quad -\mu\lambda_N \leq \lambda_T \leq \mu\lambda_N \quad \text{such that}  \quad \gamma_{T} = 0 
\end{aligned}
    \label{eq:frictionsetlaw}
\end{equation}
\noindent Therefore, the admissible values of the tangential contact force $\lambda_T$ can be represented as convex set $\mathrm{C}_T \in \{\lambda_T |-\mu\lambda_N \leq \lambda_T \leq \mu\lambda_N\}$, and slipping occurs when the relative velocity exceeds the static friction threshold. The direction of the generalized relative tangential force can be represented by $\mathrm{W_T} = [\frac{\partial \gamma_T}{\partial u}]^T$.  

\subsection{Unified Non-Smooth Dynamics}

Now using \eqref{eq:signorinicond}-\eqref{eq:frictionsetlaw}, the non-smooth impact forces can be added to \eqref{eq:model} and we get the following equations,

\begin{equation}
   \texttt{Non-Impulse Dynamics}: \quad \mathrm{M(t,q)} \dot{u} - \mathrm{h(t,u,q)} - \mathrm{W_N} \lambda_N - \mathrm{W_T} \lambda_T = 0
    \label{eqa:minimal}
\end{equation}
\begin{equation}
     \texttt{Impulse Dynamics}: \quad  \mathrm{M(t,q)} (u_E - u_A) - \mathrm{W_N} \Lambda_N - \mathrm{W_T} \Lambda_T= 0
    \label{eqa:full}
\end{equation}

\noindent The equation \eqref{eqa:minimal} represents the minimal dynamics without impulses during impacts whereas \eqref{eqa:full} represents the U-Joint dynamics with impact impulse dynamics. The velocities before and after impacts are given as $u_E$ and $u_A$. In \eqref{eqa:full}, the velocity jumps are related to impulse forces $\Lambda_N$ and $\Lambda_T$. The non-impulsive \eqref{eqa:minimal} and impulsive \eqref{eqa:full} dynamics are unified through multiplying \eqref{eqa:minimal} by Lebesgue measure $(dt)$ and \eqref{eqa:full} by Dirac atomic measure $(d\eta)$,

\begin{equation}
   \texttt{Unified Dynamics}: \  \mathrm{M(t,q)} du - \mathrm{h(t,u,q)}{dt} - \mathrm{W_N} dP_N - \mathrm{W_T} dP_T = 0
    \label{eqa:combine}
\end{equation}

In unified dynamics \eqref{eqa:combine}, the measure of velocities $du = \dot{u}dt + (u_E - u_A)d\eta$, the atomic measure for the impulse is $dP = \lambda dt + \Lambda d\eta$. For impact-free systems $d\eta = 0$. The Signorini's conditions need to be complemented with Newton's kinematical law to relate the pre- and post-impact velocities in normal and tangential directions of impact. Let $\gamma_{NA}$ and $\gamma_{TA}$ be the relative velocity before impact, and $\gamma_{NE}$ and $\gamma_{TE}$ be the relative velocity after the impact. The post-impact velocities are made slightly higher than Newton’s law estimate to capture jumps in velocities for non-impulsive contact loads.

\begin{equation}
    \begin{aligned}
        \xi_N = \gamma_{NE} + \epsilon_N \gamma_{NA} \\
        \xi_T = \gamma_{TE} + \epsilon_T \gamma_{TA}
    \end{aligned}
\end{equation}
Through unilateral primitive inclusions \eqref{eq:uprimpulse}, impulse forces can be represented by the convex set of all admissible impulse forces without violating \eqref{eq:signorinicond}, 
\begin{equation}
    \begin{aligned}
        -dP_N \in \mathrm{Upr(\xi_N)} \\
        -dP_T \in \mu dP_N \mathrm{Sgn(\xi_T)}
    \end{aligned}
    \label{eq:uprimpulse}
\end{equation}
For more details on above formulations for non-smooth systems refer to \cite{Glocker2001}.

\subsection{Numerical Simulation Algorithm}

In this study, we employ a time-stepping integration scheme for non-smooth dynamical systems \eqref{eqa:combine}, following the approach of \cite{Flores2010} with Lemke's solver \cite{Lemke1965}. This method is particularly effective for simulating unilateral rigid impacts with frictional contacts, as found in U-Joints with clearance. Unlike conventional ODE solvers, which assume smooth system behavior, time-stepping methods inherently accommodate the velocity discontinuities and stick-slip transitions characteristic of impact dynamics, eliminating the need for error-prone event detection. Recent studies \cite{Golembiewski2024}, demonstrate that time-stepping integrators offer superior robustness and accuracy in capturing these complex interactions, reinforcing their suitability for rigid body contact problems. \par
 
 Consider the generalized coordinates and velocities before and after impact are $\{q_A, u_A\}$ at time $t_A$ and $\{q_E, u_E\}$ at time $t_E$. The subscript $A$ represents before impact and $E$ represents after impact. The algorithm calculates generalized coordinates at midpoint time $q_M = q_A + \frac{1}{2}\mathrm{\Delta} t u_A$. At this midpoint, find out all permissible contact set $\mathrm{h_M}$. The set of all closed contacts is defined as \eqref{eqa:closedcontactset}. Likewise, for every $i \in \mathrm{H_M}$ the linear complementary problem \eqref{eq:LCP} is solved for post-impact velocities and impulse forces in normal $(P_N)$ and tangential directions $(P_R)$. The total impulse contact force is $P_L$, \par
\begin{equation}
\begin{aligned}
\begin{pmatrix}
\xi_N \\
\xi_R \\
P_L
\end{pmatrix} = 
\begin{pmatrix}
\mathrm{W}_{NM}^T \mathrm{M}_M^{-1} \mathrm{W}_{NM} (\mathrm{W}_{NM} - \mathrm{W}_{TM} \mu) & \mathrm{W}_{NM}^T \mathrm{M}_M^{-1} \mathrm{W}_{TM} & 0 \\
\mathrm{W}_{TM}^T \mathrm{M}_M^{-1} \mathrm{W}_{NM} (\mathrm{W}_{NM} - \mathrm{W}_{TM} \mu) & \mathrm{W}_{TM}^T \mathrm{M}_M^{-1} \mathrm{W}_{TM} & \mathrm{I} \\
2\mu & -\mathrm{I} & 0
\end{pmatrix}
\begin{pmatrix}
P_N \\
P_R \\
\xi_L
\end{pmatrix} 
\\ + 
\begin{pmatrix}
\mathrm{W}_{NM}^T \mathrm{M}_M^{-1} \mathrm{h}_M \mathrm{\Delta} t + (\mathrm{I} + \varepsilon_N) \gamma_{NA} \\
\mathrm{W}_{TM}^T \mathrm{M}_M^{-1} \mathrm{h}_M \mathrm{\Delta} t + (\mathrm{I} + \varepsilon_T) \gamma_{TA} \\
0
\end{pmatrix}
\end{aligned}
\label{eq:LCP}
\end{equation}

At each time-step the post-impact positions  and velocities are updated using \eqref{eq:postimpact_velo},
\begin{equation}
\begin{aligned}
        u_E = u_A + \mathrm{M_M}^{-1} \mathrm{h_M}\mathrm{\Delta} t + \mathrm{M_M}^{-1}(\mathrm{W_{NM}} - \mu \mathrm{W_{TM}})P_{N} + \mathrm{M_M}^{-1}\mathrm{W_{TM}}P_{R}
        \\
        q_E = q_M + \frac{1}{2}\mathrm{\Delta} t u_E
\end{aligned}
    \label{eq:postimpact_velo}
\end{equation}
where $\mathrm{M_M}$, $\mathrm{h_M}$, $\mathrm{W_{NM}}$ and $\mathrm{W_{TM}}$ are system matrices evaluated at the midpoint time and $I$ is identity matrix. The numerical simulation time step was kept $10^{-5}$ to capture the post-impact bounces accurately. This step size was chosen as a trade-off between computational efficiency and numerical accuracy, ensuring a balance between solution precision and feasible simulation time. 

\begin{table}[ht]
    \centering
    \caption{Numerical simulation parameter values.}
    \renewcommand{\arraystretch}{1.2}
    \resizebox{\textwidth}{!}{
    \begin{tabular}{ccccc}
        \hline
        \textbf{Parameter(s)} & \textbf{Description} & \textbf{Value} & \textbf{Unit} \\
        \hline
         $\mathrm{{J_1}}$ & Inertia of input shaft & 0.014 & $\text{kg m}^2$ \\
         $\mathrm{J_{2x}}$, $\mathrm{J_{2y}}$, $\mathrm{J_{2z}}$ & Inertia of crosspiece along rotational axis & 0.00111, 0.00202, 0.00111 & $\text{kg m}^2$ \\
         $\mathrm{{J_3}}$ & Inertia of output shaft & 0.012 & $\text{kg m}^2$ \\
        $\mathrm{{K_s}}$ & Torsional stiffness of output shaft& $1000 $ & $\text{Nm/rad}$ \\
         $\mathrm{\mathrm{C_s}}$ &Damping coefficient & 5 & $\text{Nm·s/rad}$ \\
        $\mathrm{R_1}$ & Radius of crosspiece arm cap& 0.02 & m \\
        $\mathrm{C}$ & Radial Clearance & 50 & $\mathrm{\mu}$m \\
       $\mathrm{\mathrm{\beta}}$ & U-Joint angle& 5 & degrees \\
       $\mathrm{L}$ &Crosspiece arm length& 0.04 & m \\
        $\mathrm{\epsilon_N}$, $\epsilon_T$ & Coefficient of restitution in normal and tangential direction& 0.45, 0.45 & - \\
       $\mathrm{\mu}$ & Coefficient of friction & 0.8 & - \\
       $\mathrm{\Omega}$ & Forcing Frequency & 100 & \text{rad/sec} \\
       $\mathrm{T_0}$ & Forcing Amplitude & 1.0 & \text{Nm} \\
    \hline
    \end{tabular}
    }
    \label{tab:my_label}
\end{table}

\section{Result and Discussion} \label{results}

\begin{figure}[ht]
    \centering
  
    \subfigure[]{\includegraphics[width=0.24\linewidth]{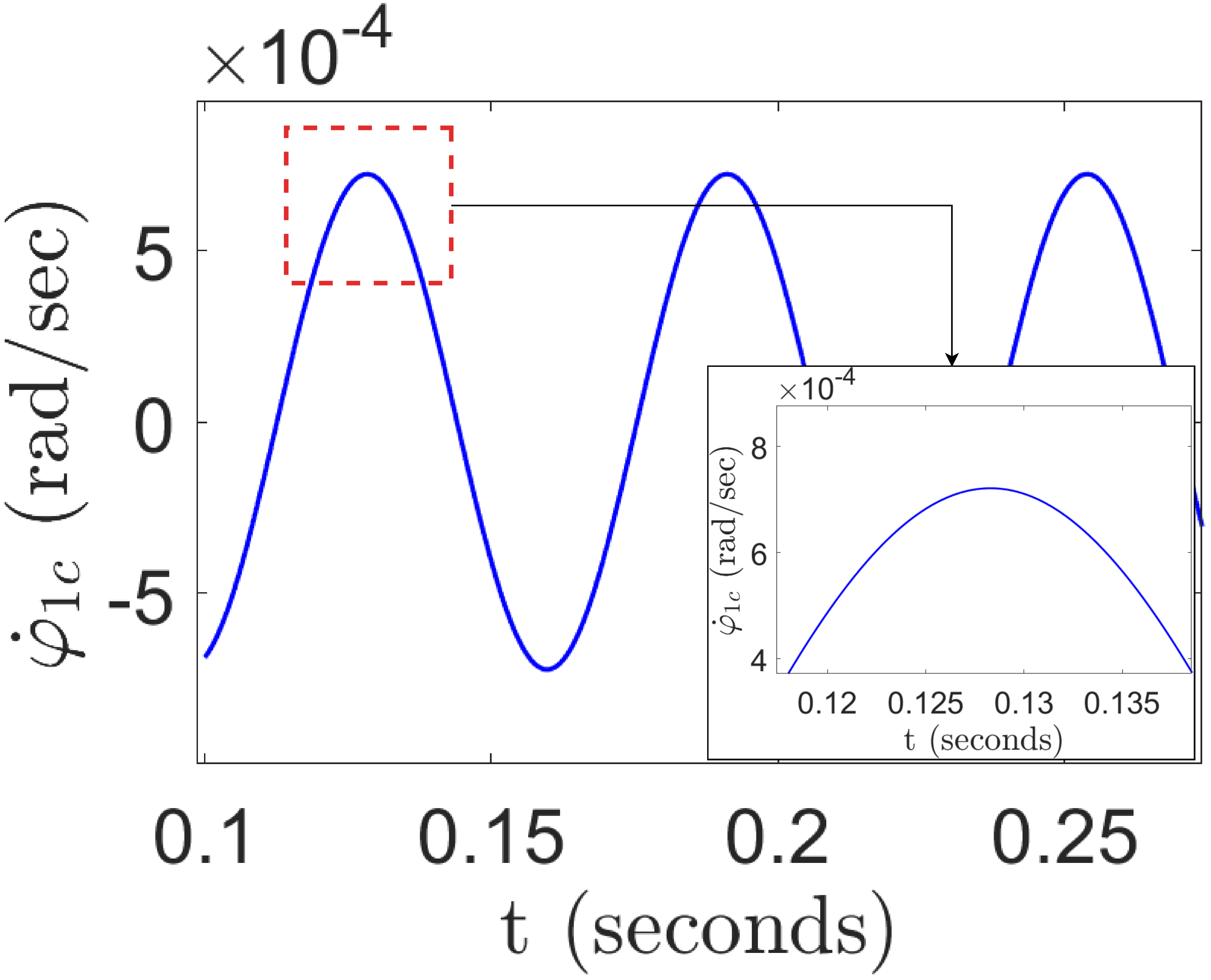}}
    \subfigure[]{\includegraphics[width=0.23\linewidth]{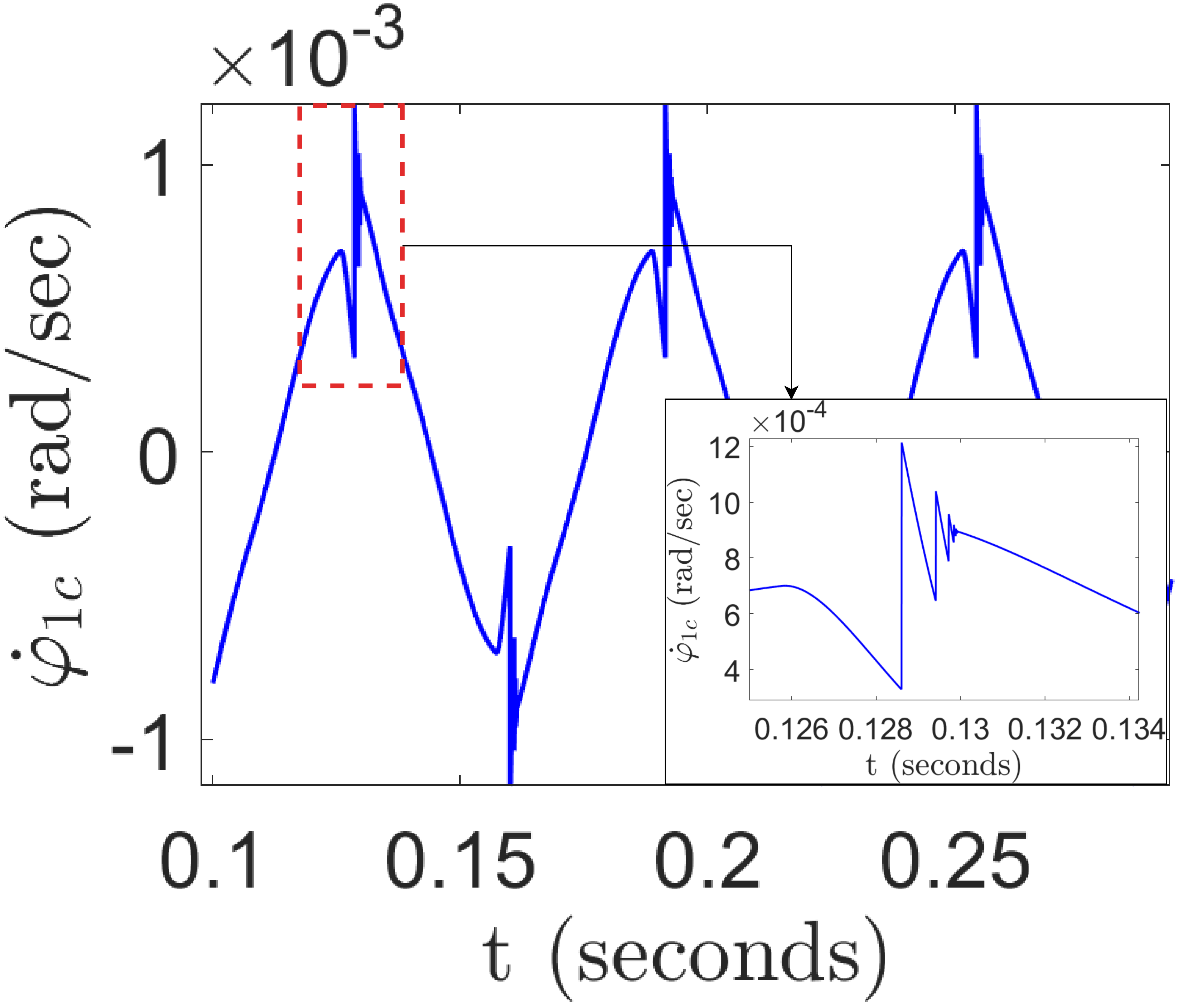}}
    \subfigure[]{\includegraphics[width=0.23\linewidth]{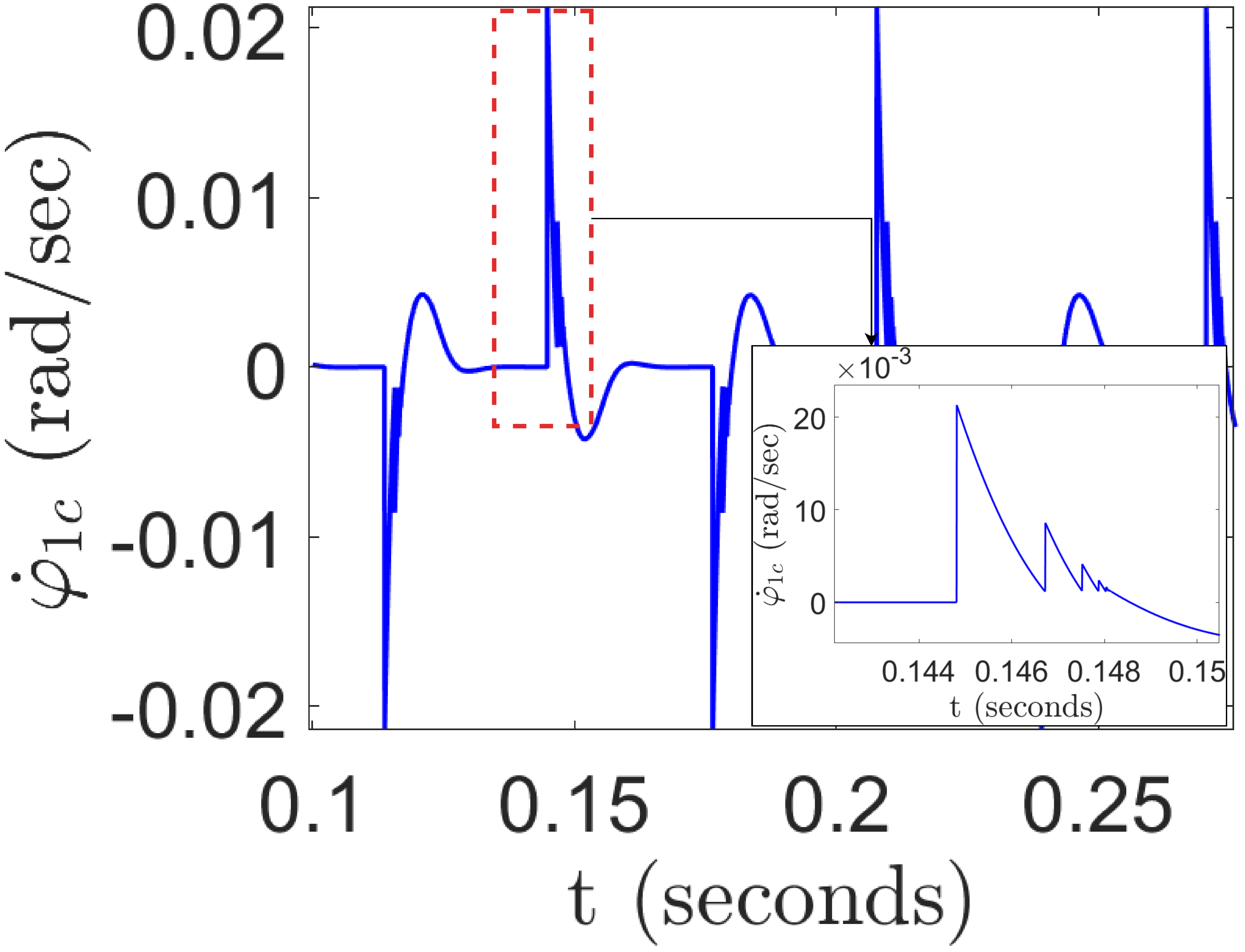}}
    \subfigure[]{\includegraphics[width=0.24\linewidth]{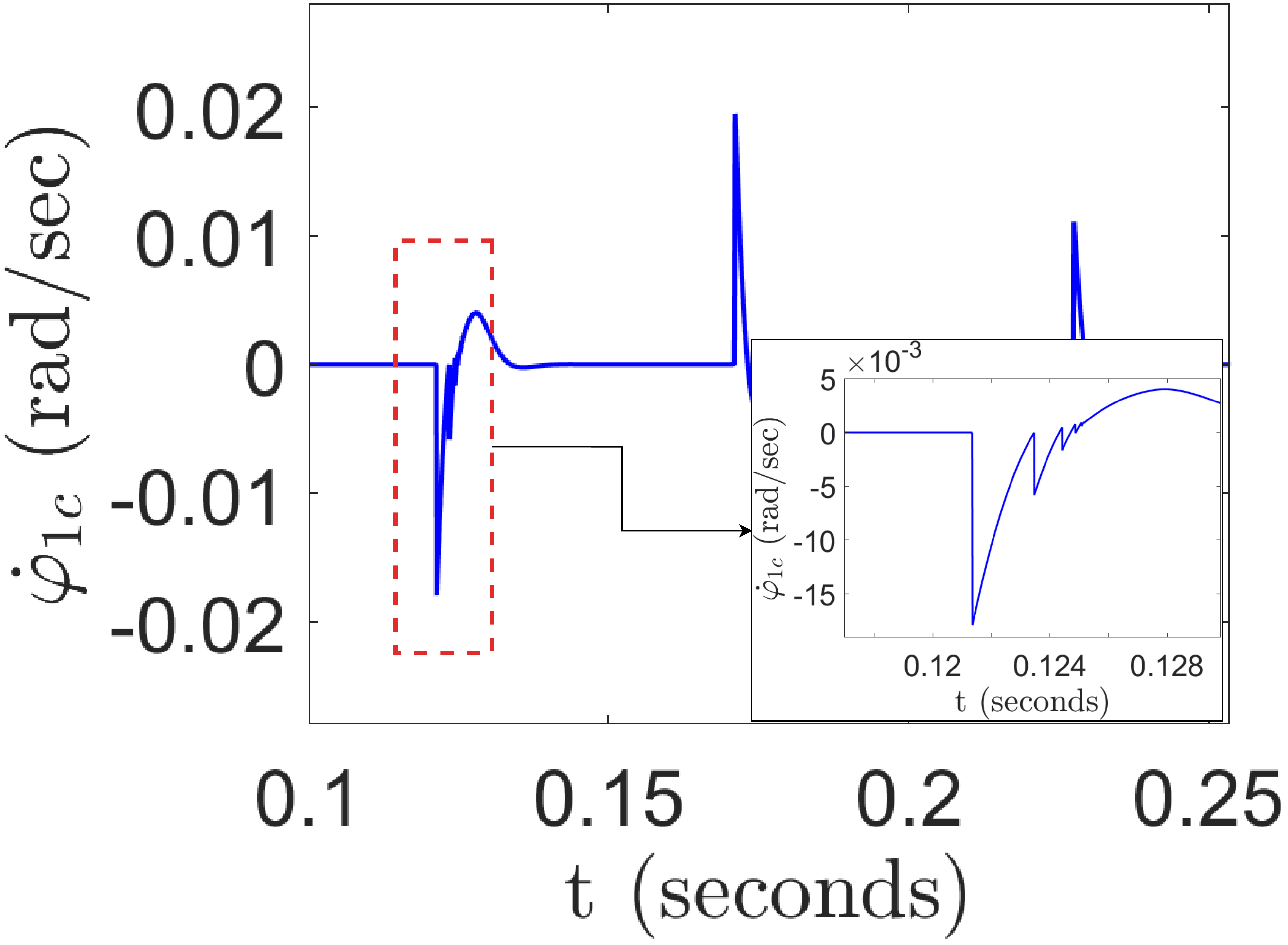}}

   \caption{Numerical simulation time response for clearance size (a) \(C = 0\) \(\mu\)m, (b) \(C = 0.05\) \(\mu\)m, (c) \(C = 10\) \(\mu\)m, and (d) \(C = 50\) \(\mu\)m.}

    \label{fig:results_time_response}
\end{figure}

\begin{figure}[ht]
    \centering
        
    % Phase Space
    \subfigure[]{\includegraphics[width=0.24\linewidth]{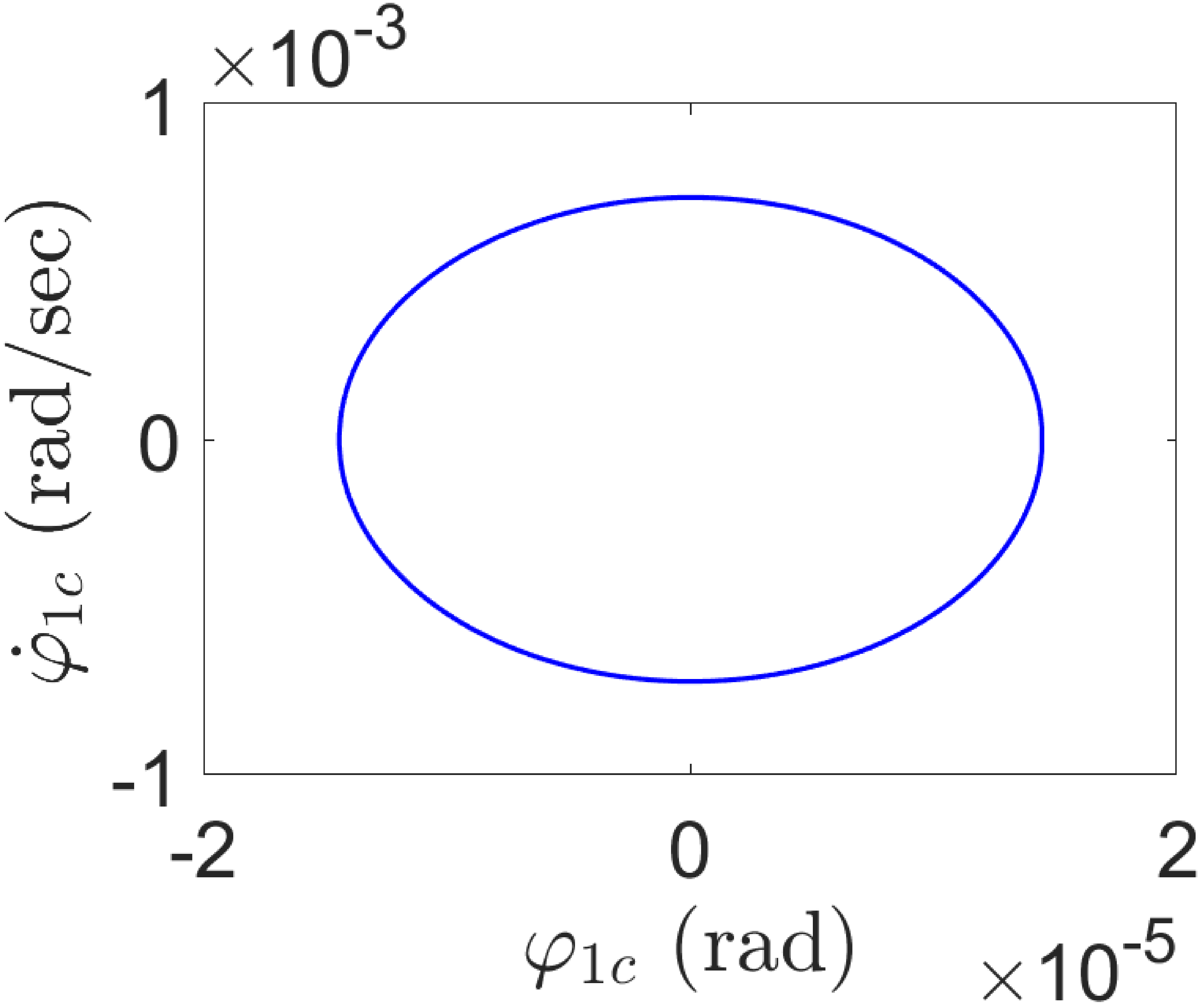}}
    \subfigure[]{\includegraphics[width=0.24\linewidth]{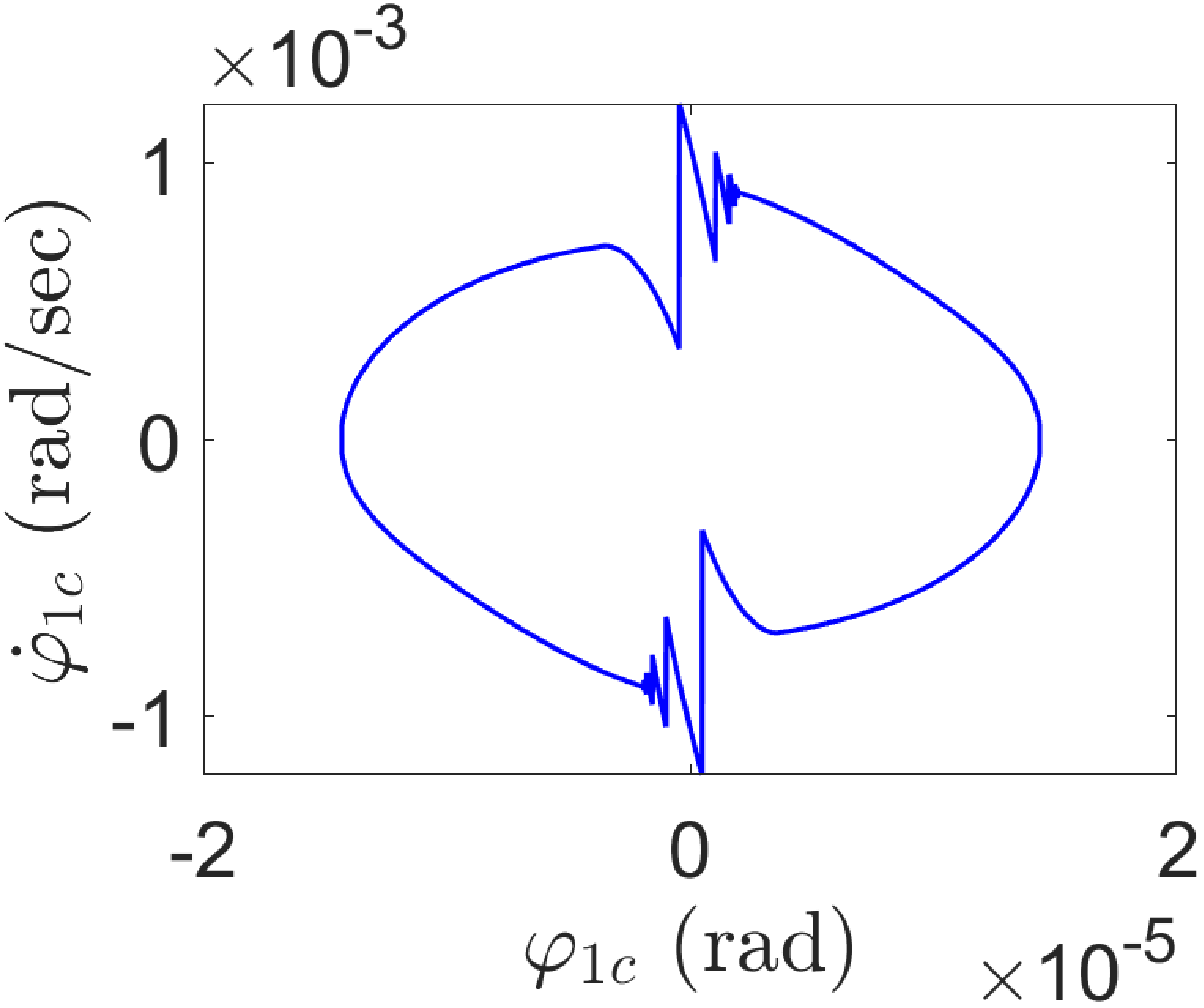}}
    \subfigure[]{\includegraphics[width=0.24\linewidth]{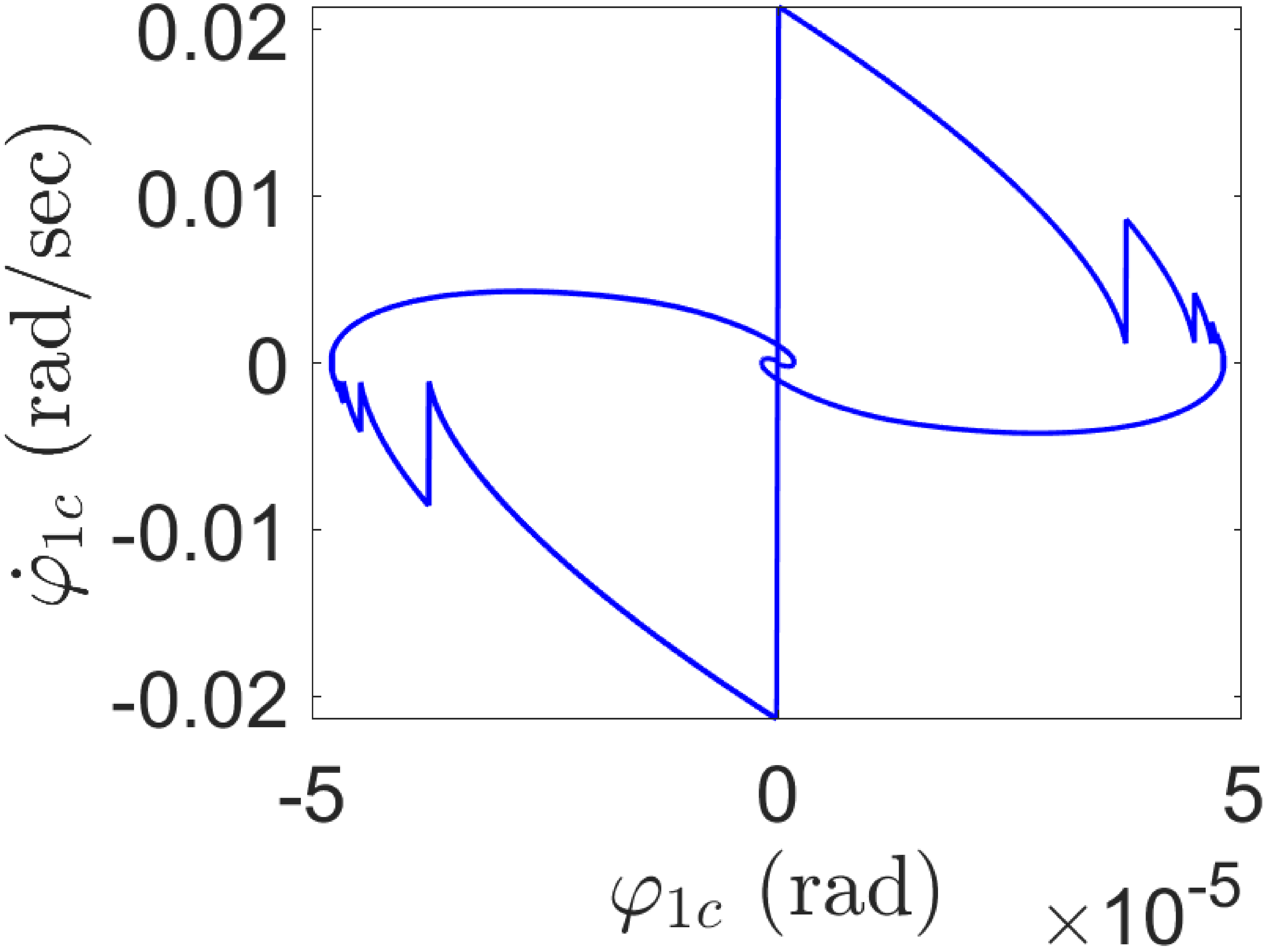}}
    \subfigure[]{\includegraphics[width=0.24\linewidth]{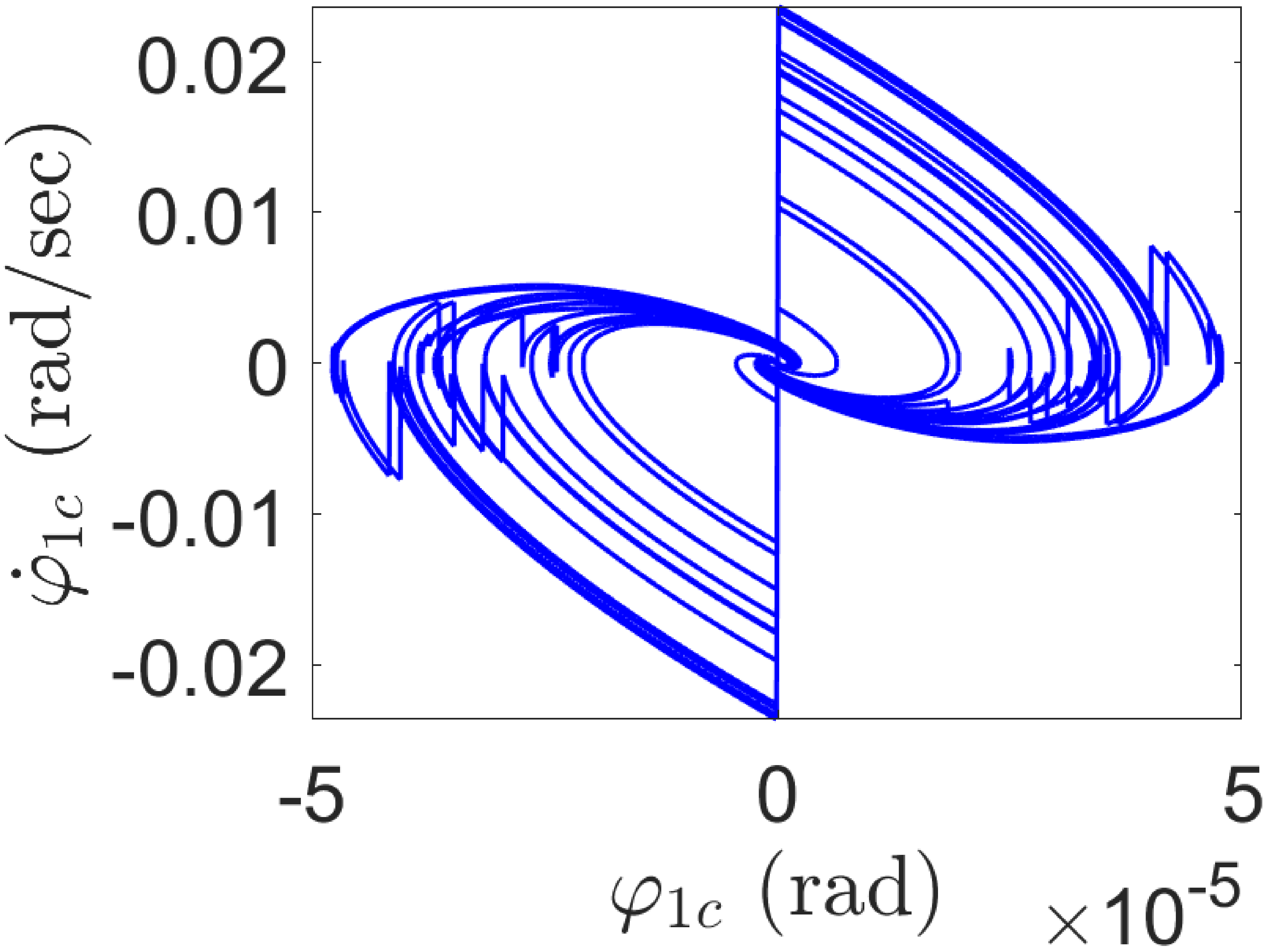}}
    
\caption{Numerical simulation phase portrait of \((\varphi_{1c}, \dot{\varphi}_{1c})\) for clearance size (a) \(C = 0\) \(\mu\)m, (b) \(C = 0.05\) \(\mu\)m, (c) \(C = 10\) \(\mu\)m, and (d) \(C = 50\) \(\mu\)m.}

    \label{fig:results_phase_space}
\end{figure}

\begin{figure}[ht]
    \centering
    
    % Phase Space
    \subfigure[]{\includegraphics[width=0.24\linewidth]{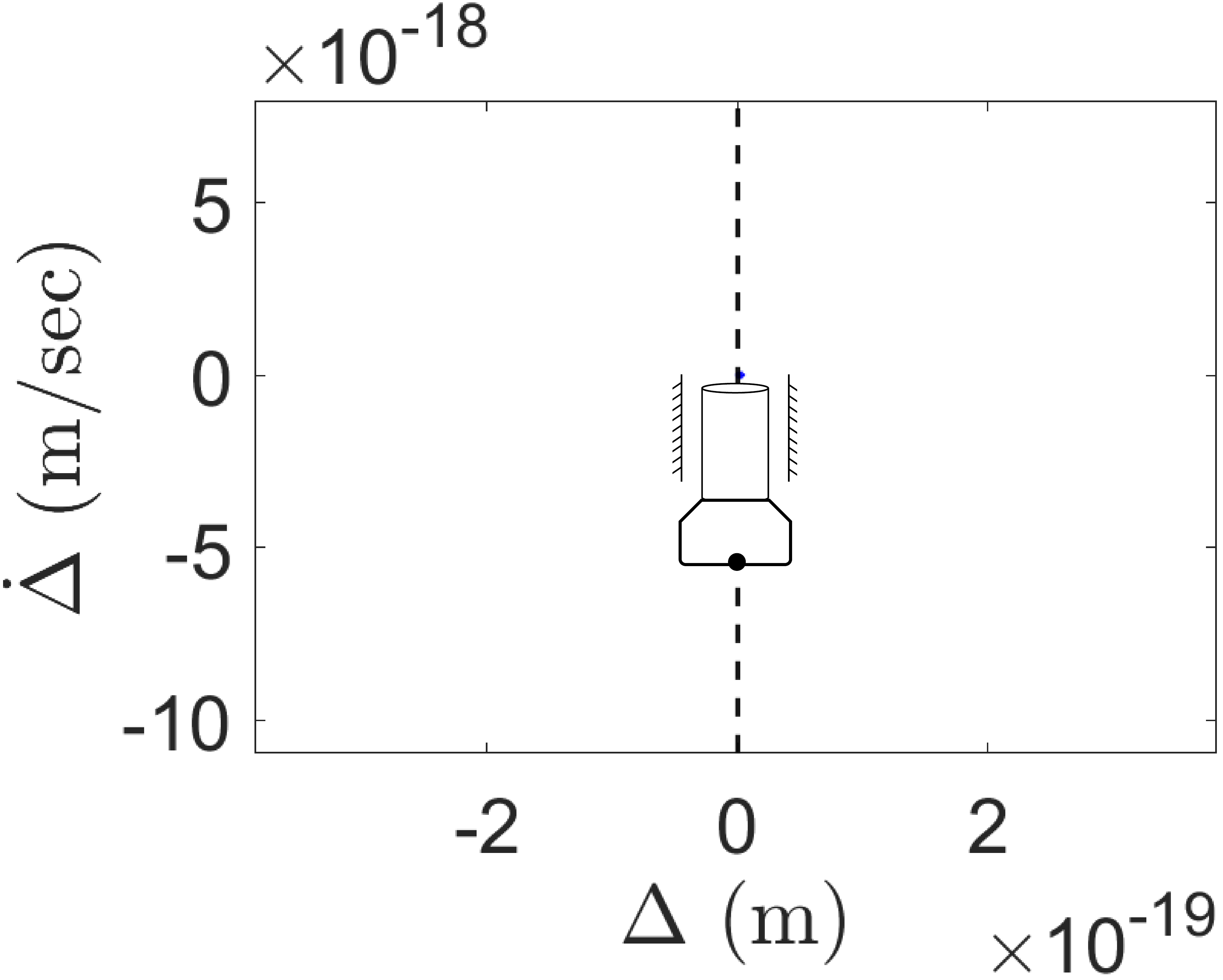}}
    \subfigure[]{\includegraphics[width=0.24\linewidth]{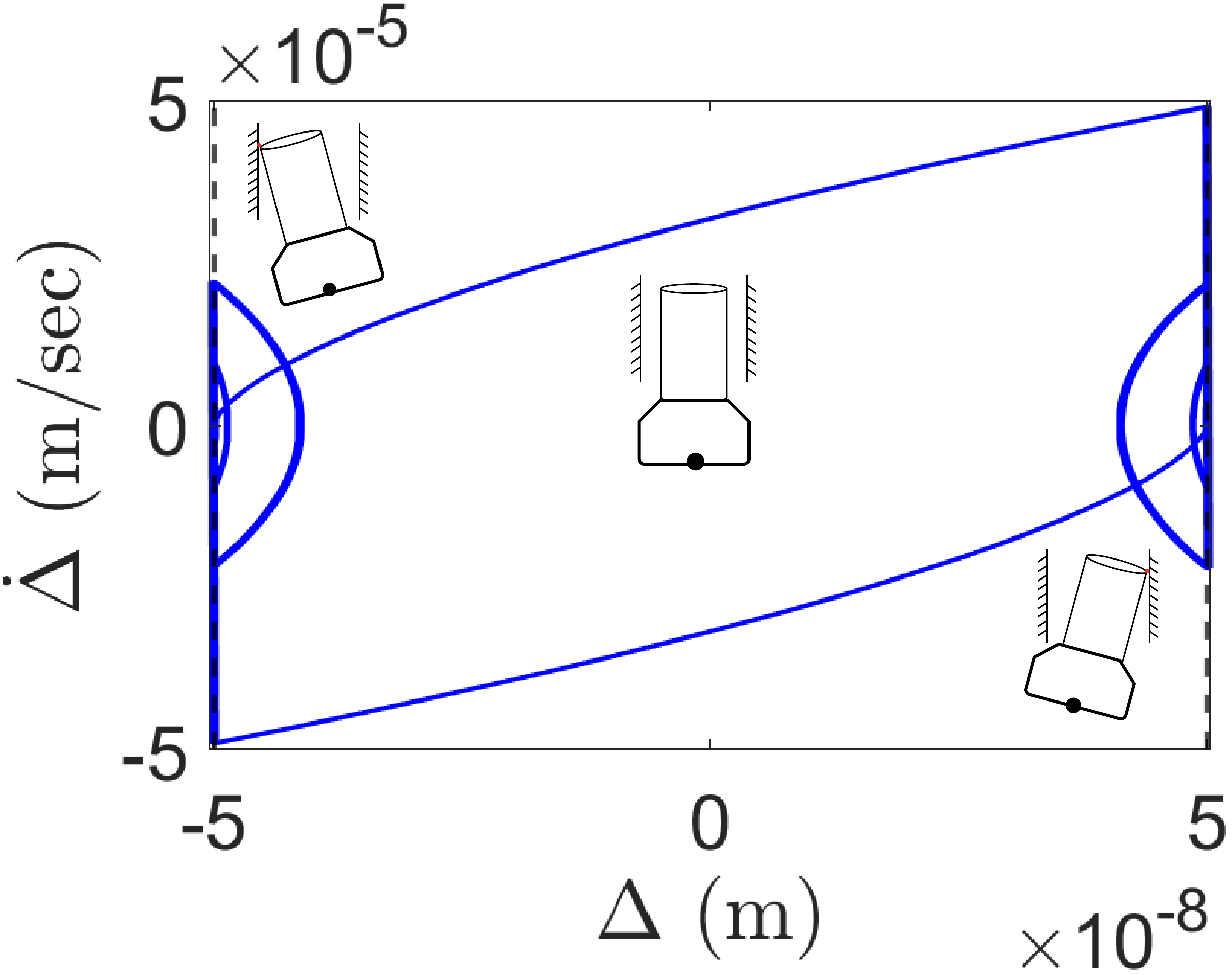}}
    \subfigure[]{\includegraphics[width=0.23\linewidth]{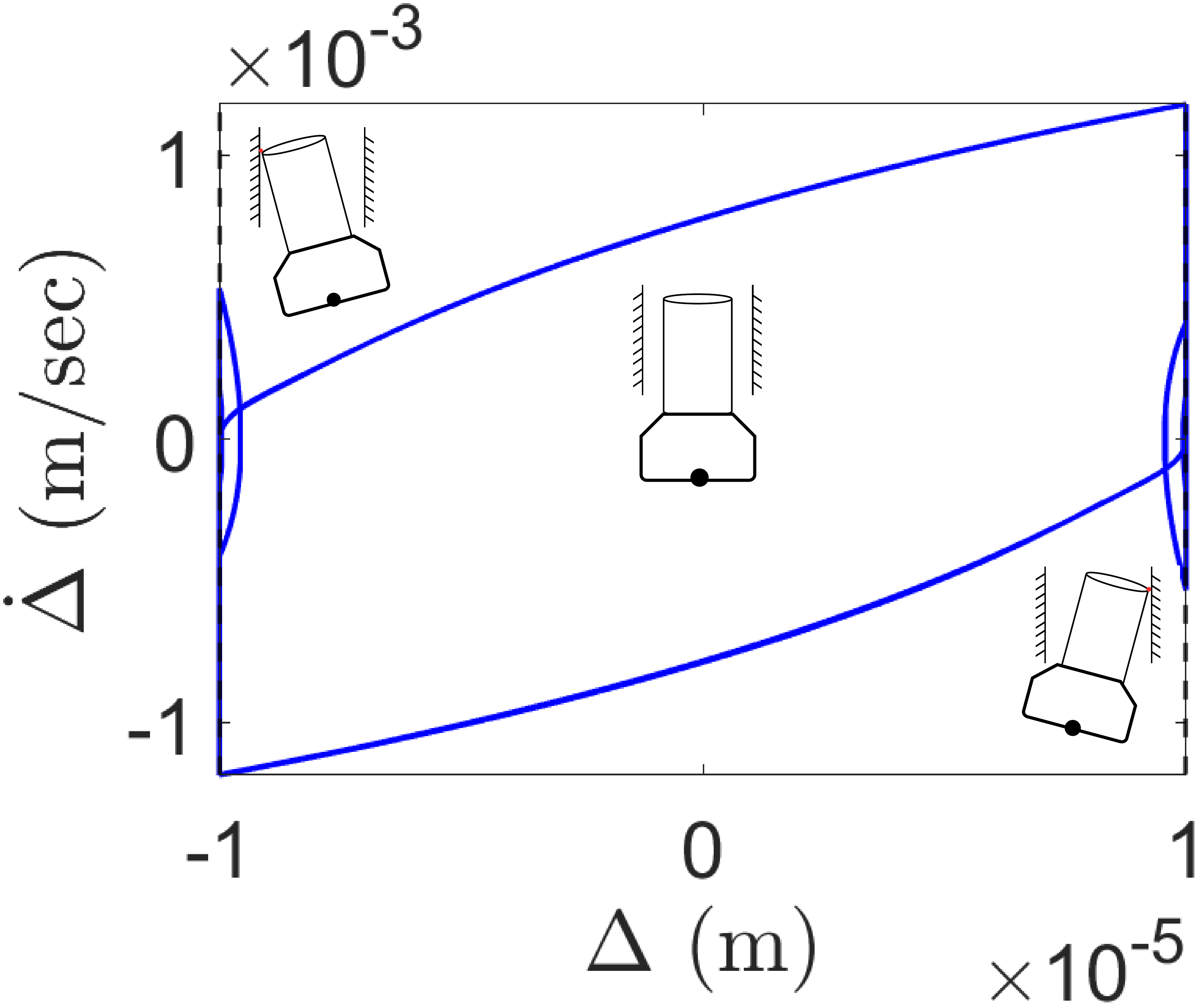}}
    \subfigure[]{\includegraphics[width=0.24\linewidth]{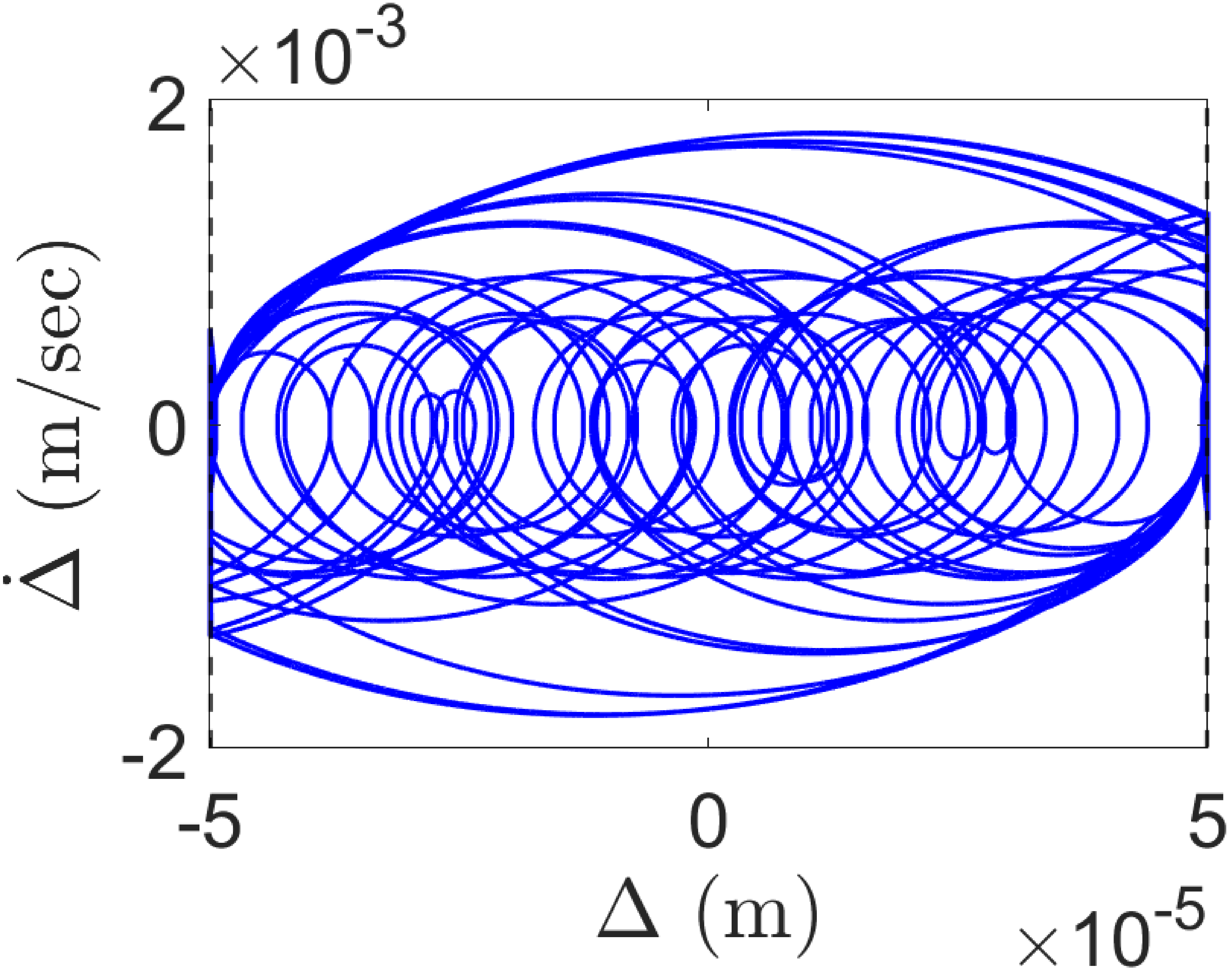}}
 \caption{Numerical simulation phase portrait of \((\mathrm{\Delta}, \mathrm{\dot{\Delta}})\) for clearance size (a) \(C = 0\) \(\mu\)m, (b) \(C = 0.05\) \(\mu\)m, (c) \(C = 10\) \(\mu\)m, and (d) \(C = 50\) \(\mu\)m.}

    \label{fig:results_gap_function_phase}
\end{figure}

The numerical simulations, based on the unified non-smooth dynamics framework in \eqref{eqa:combine} with parameters from Table \ref{tab:my_label}, were initialized at $q_A = [0, 0]^T$ and $u_A = [0, 0]^T$. Figures \ref{fig:results_time_response} (a) - (d) present the steady-state time response for varying clearances $\mathrm{C}$. For $\mathrm{C} = 0$, the system exhibits stable period-1 oscillations with no impacts. As clearance increases to $\mathrm{C} = 0.05 \, \mu\mathrm{m}$, velocity discontinuities emerge due to wall collisions, introducing high-frequency oscillations. At $\mathrm{C} = 10 \, \mu\mathrm{m}$, these impacts intensify, producing sharp velocity jumps. With further increase to $\mathrm{C} = 50 \, \mu\mathrm{m}$, the system enters a chaotic regime, characterized by erratic impacts, large velocity fluctuations, and complex contact dynamics, potentially leading to operational instability and excessive wear.

Figures \ref{fig:results_phase_space} (a) - (d) show the 2D steady-state phase portraits in the $(\varphi_{1c}, \dot{\varphi}_{1c})$ plane. For zero clearance, the motion is purely periodic. At $\mathrm{C} = 0.05 \, \mu\mathrm{m}$, the phase trajectories exhibit symmetric impacts at $\varphi_{1c} = 0$ with brief free flight intervals. At $\mathrm{C} = 10 \, \mu\mathrm{m}$, the trajectories sharply reverse at the boundaries, consistent with rigid contact. For large clearance $\mathrm{C} = 50 \, \mu\mathrm{m}$, the phase space exhibits concentric, spiraling loops, indicating a transition to quasi-periodic or chaotic motion, with small asymmetries due to nonlinear friction effects.

The gap function phase analysis in Figures \ref{fig:results_gap_function_phase} (a) - (d) captures the crosspiece motion within the yoke. For $\mathrm{C} = 0$, the gap remains closed, reflecting complete motion transmission. At $\mathrm{C} = 0.05 \, \mu\mathrm{m}$, high-frequency micro-impacts occur near the boundaries, forming enclosed loops due to repeated collisions, followed by sticking on the wall due to friction. With increasing clearance, the free-flight duration extends as crosspiece needs to traverse longer distance before collision, reducing confinement within the yoke, while the dense web of overlapping loops at $\mathrm{C} = 50 \, \mu\mathrm{m}$ reflects chaotic, irregular motion with frequent, rapid contacts. Impact-induced vibrations are common in rotating systems with clearances, but chaotic or chattering impacts are especially harmful, causing noise, wear, fatigue, and potential structural failure.
 
\section{Conclusion} \label{conclu}

This study develops a multibody model for U-Joint impact dynamics with radial clearance between crosspiece and input yoke, using a rigid unilateral contact law and computationally efficient time-stepping algorithm to capture interactions between the crosspiece and yoke. Unlike prior models, it includes crosspiece inertia and friction between crosspiece and yoke, improving dynamic accuracy. Phase plane analysis of the gap function reveals that larger clearances lead to chaotic, high-energy impacts linked to wear and noise. The model is applicable to automotive, aerospace, and medical systems where precision motion is critical. Future work will extend the framework to multi-shaft drivelines and experimental validation of the theoretical findings presented in this paper.

\bibliographystyle{splncs03_unsrt} %titles are missing 
\bibliography{author}
\end{document}